\documentclass[preprint]{revtex4-2}
\usepackage[utf8]{inputenc}
\usepackage{hyperref}
\usepackage{amsmath}
\usepackage{bm}
\usepackage[version=4]{mhchem}
\usepackage{caption}
\usepackage{graphicx}
\usepackage{subcaption}
\usepackage{float}
\usepackage{comment}
\usepackage[export]{adjustbox}
\usepackage{tabularx}
\usepackage{xcolor}
\usepackage{cleveref}
\usepackage{color}
\definecolor{shiningblue}{rgb}{0.3,0.68,0.89}

\newcommand\at[2]{\left.#1\right|_{#2}}

\begin{document}
\title{Mechanically-driven growth and competition in a Voronoi model of tissues}
\author{Louis Br\'ezin}
\email{lbrezin@bu.edu}
\author{Kirill S. Korolev}
\email{korolev@bu.edu}
\affiliation{Department of Physics, Graduate Program in Bioinformatics and Biological Design Center,
Boston University, Boston, Massachusetts 02215, USA}
\date{\today}

\begin{abstract}
    The mechanisms leading cells to acquire a fitness advantage and establish themselves in a population are paramount to understanding the development and growth of cancer.
    Although there are many works that study separately either the evolutionary dynamics or the mechanics of cancer, little has been done to couple evolutionary dynamics to mechanics. To address this question, we study a confluent model of tissue using a Self-Propelled Voronoi (SPV) model with stochastic growth rates that depend on the mechanical variables of the system. 
    The SPV model is an out-of-equilibrium model of tissue derived from an energy functional that has a jamming/unjamming transition between solid-like and liquid-like states.
    By considering several scenarios of mutants invading a resident population in both phases, we determine the range of parameters that confer a fitness advantage and show that the preferred area and perimeter are the most relevant ones. We find that the liquid-like state is more resistant to invasion and show that the outcome of the competition can be determined from the simulation of a non-growing mixture.
    Moreover, a mean-field approximation can accurately predict the fate of a mutation affecting mechanical properties of a cell. Our results can be used to infer evolutionary dynamics from tissue images, understand cancer-suppressing effects of tissue mechanics, and even search for mechanics-based therapies.
\end{abstract}

\maketitle

\section{Introduction}

Tissues are communities of cells and undergo evolutionary processes.
Cell competition, in which one cell type takes over another, is particularly relevant for cancer invasion, but also in developmental processes \cite{tamori_cell_2011,morata_minutes_1975,korolev_turning_2014}.
Cell competition between a mutant and a resident population is the result of changes that alter the growth rate of the mutants with respect to the resident. 
If a mutant acquires a higher growth rate, it will eventually fixate and take over the entire population.

In normal tissues, there is a turnover of cells over time \cite{sender_distribution_2021}: cells undergo division and are eliminated by apoptosis or cell extrusion \cite{guillot_mechanics_2013}.
The balance between cell division and extrusion controls the overall growth of the tissue. 
In what follows, we will refer to cell extrusion without distinction between apoptosis and extrusion because, at our level of description, they both contribute to the removal of a cell from the monolayer. 
We are interested in this paper in the mechanical regulation of growth and its implications in cell competition \cite{tamori_cell_2011,matamoro-vidal_multiple_2019,levayer_solid_2020}. 
At the cellular level, there are mechanotransductive proteins that provide mechanical feedback to cells, whose levels can promote or hinder growth \cite{wagstaff_mechanical_2016,aragona_mechanical_2013,balasubramaniam_investigating_2021}.
On a macroscopic level, it has been shown that exerting pressure on cell spheroids hinders growth, both by lowering the division rate and by increasing the rate of apoptosis \cite{montel_stress_2011}.
In cell monolayers, cells try to maintain a constant homeostatic density by extruding cells in denser areas, where the pressure is greater \cite{eisenhoffer_crowding_2012}.
The idea that cell growth is regulated by pressure is not new \cite{shraiman_mechanical_2005,basan_homeostatic_2009}, but to our knowledge its implication for cell competition has not been explored outside of continuum models \cite{ranft_mechanically_2014}.

To investigate the effect of mechanics on cell competition, we study a commonly used model of a confluent cell monolayer: the Self-Propelled-Voronoi (SPV) model. We supplement it with a stochastic growth process with birth and death rates controlled by tissue mechanics. In the SPV model, the motion of the cell centers is derived from an energy functional, and the network of cells is constructed using a Voronoi tessellation. The energy functional is the same as in the Vertex Model~\cite{farhadifar_influence_2007,alt_vertex_2017}, but we model the motion of the cell centers rather than the vertices. An important aspect of both models is that they predict a jamming-unjamming transition~\cite{bi_density-independent_2015,bi_motility-driven_2016,chiang_glass_2016} that has been suggested to play a central role in several biological processes~\cite{mitchel_primary_2020,park_unjamming_2015,atia_are_2021}.

After introducing the SPV model for a heterogeneous population composed of two types of cells and its birth-death dynamics, we simulate invasion scenarios with all possible single-parameter mutants to determine the outcome of the competition. We find that the fate of a mutant can be predicted from a simulation of a mutant-ancestor mixture without cell division or death. Using this fact and a mean-field approximation, we explain the results of our simulation by a simple argument based on energy minimization. 
    
\section{Model}
    \begin{figure}
        \centering
        \includegraphics[width=.7\textwidth]{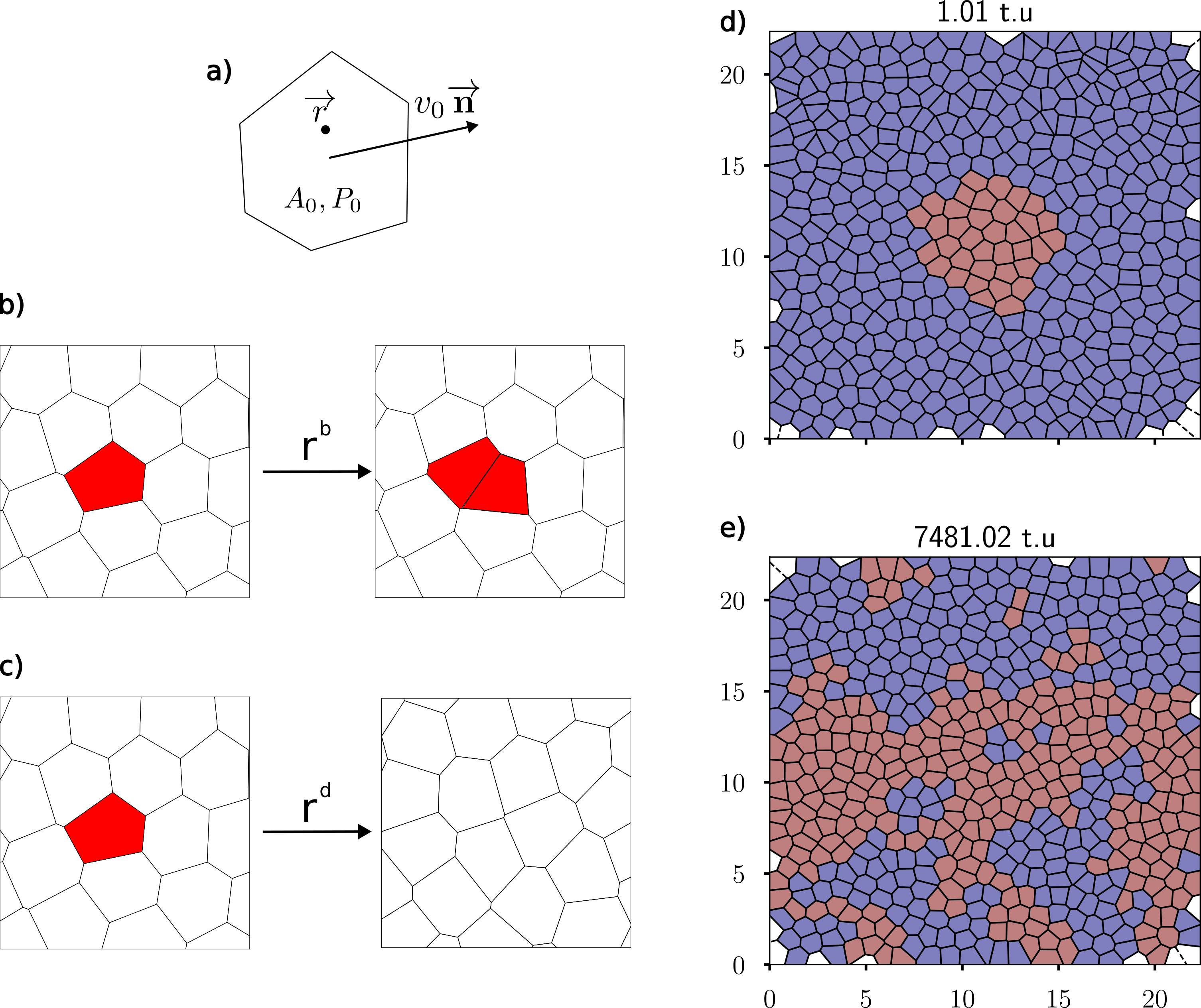}
        \caption{Overview of model. a) A Voronoi cell and its parameters. b) Example of a cell division. The red cell is split in half by a segment that goes through the center of the cell, with the angle of the segment chosen at random. Two new cell centers are created at a certain distance normal to this segment, and a new Voronoi tessellation is computed. c) Example of cell death/extrusion in which a cell center is removed and a new Voronoi tessellation is created. d) Initial and e) final configuration of an invasion simulation, where a small droplet of mutant with a lower preferred area is introduced with fraction $f=0.1$. A time unit represents 100 simulation time steps.}
        \label{fig:overview}
    \end{figure}
    \subsection{Heterogeneous Self-Propelled Voronoi model}
    We use the Self-Propelled Voronoi (SPV) model to describe the dynamics of a confluent monolayer of active cells, as found in tissues \cite{bi_motility-driven_2016}. 
    In this model, the cell centers evolve according to equations of motions derived from an energy functional, with self-propulsion motility. 
    Once the positions of the cell centers are computed, the shapes of the cells are determined using a Voronoi tessellation, i.e. by maximizing the size of each cell with respect to the position of its cell center.
    Other popular models such as the Vertex Model \cite{farhadifar_influence_2007} or the Cellular Potts model \cite{graner_simulation_1992} use more degrees of freedom to compute the shape of cells, but the simple aspect of having the cells themselves as degrees of freedom make evolutionary dynamics much easier to track, especially since we have little interest in the details of the cell shapes.
    
    The energy functional for the cells is the same regardless of the chosen model:
    \begin{equation}
        E = \sum_\alpha \frac{K_\alpha}{2} (A_\alpha - A_\alpha^0)^2 + \sum_{<i,j>} \Lambda_{ij} l_{ij} + \sum_\alpha \frac{\Gamma_\alpha}{2} P_\alpha^2,
        \label{eq_energy_raw}
    \end{equation}
    where the index $\alpha$ represents a cell in the tissue.
    Although the requirement for the cells to form a Voronoi tessellation appears as an added constraint compared to the Vertex Model, the dynamics of the network of cells is very similar in both cases far from the low noise limit \cite{sussman_anomalous_2018,sussman_no_2018}. 
    
    We are interested in the competition between two types of cells, a resident population labeled with a subscript $_r$ and a mutant population labeled with a subscript $_m$. 
    We assume for now that cells are homogeneous within each type. 
    We define the total number of cells $N = N_r + N_m$, and the total energy of the population of cells is: 
    \begin{equation}
        E = \frac{K_r}{2}\sum_{\alpha=1}^{N_r}  (A_\alpha - A_r^0)^2 + \frac{\Gamma_r}{2}\sum_{\alpha=1}^{N_r}  (P_\alpha - P_r^0)^2
        + \frac{K_m}{2}\sum_{\alpha=N_r+1}^{N}  (A_\alpha - A_m^0)^2 + \frac{\Gamma_m}{2}\sum_{\alpha=N_r+1}^{N}  (P_\alpha - P_m^0)^2
    \end{equation}
    A common choice is to normalize all lengths by $\sqrt{A_r^0}$, fixing the normalized preferred area to unity. However, because our simulation using cellGPU \cite{sussman_cellgpu_2017} is setup to have a fixed volume $L^2$ set by the initial number of cells, we believe it is appropriate to normalize the lengths by $L/\sqrt{N_0}$, where $L$ is the length of the simulation box and $N_0$ is the initial number of cells. 
    With this normalization, the average cell area is determined by the cell number at time $t$:
    \begin{equation}
        \langle a \rangle = \frac{1}{N(t)}\sum_{\alpha=1}^{N(t)} a_\alpha = \frac{N_0}{N(t)}
        \label{eq:volumeConservation}
    \end{equation}
    
    The energy is normalized by $K_r L^4/N_0^2$. The dimensionless total energy $e$ reads
    \begin{equation}
        e = \frac{1}{2}\sum_{\alpha=1}^{N_r}  (a_\alpha - a_r^0)^2 + \frac{\bar{\Gamma}_r}{2}\sum_{\alpha=1}^{N_r}  (p_\alpha - p_r^0)^2
        + \frac{\bar{K}}{2}\sum_{\alpha=N_r+1}^{N}  (a_\alpha - a_m^0)^2 + \frac{\bar{\Gamma}_m}{2}\sum_{\alpha=N_r+1}^{N}  (p_\alpha - p_m^0)^2
        \label{eq_energy_dimless}
    \end{equation}
    There are two dimensionless variables, the areas $a_\alpha = A_\alpha N_0 /L^2$ and the perimeters $p_\alpha = P_\alpha \sqrt{N_0}/L$, and 7 dimensionless parameters.
    Among those 7 parameters, four are so-called geometrical parameters $a_r^0,a_m^0,
    p_r^0, p_m^0$ and three are mechanical parameters $\bar{K}, \bar{\Gamma}_r,\bar{\Gamma}_m$.
    
    At each timestep, the force exerted on cell center $i$ is given by
    \begin{equation}
        \bm{F}_i = -\frac{\partial E}{\partial \bm{r}_i} = -\sum_{j \in \text{NN}(i)} \frac{\partial E_j}{\partial \bm{r}_i} - \frac{\partial E_i}{\partial \bm{r}_i},
        \label{eq_force_center}
    \end{equation}
    where NN$(i)$ refers to the nearest neighbors of cell $i$. The equation of motion for cell center $i$ is
    \begin{align}
        \frac{\partial \bm{r}_i}{\partial t} = \mu \bm{F}_i + v^0 \bm{n}_i \label{eq_motion_cellcenter}
    \end{align}
    with $\mu$ the mobility of cells taken to be homogeneous amongst cells, and $\bm{n}_i = (\cos \theta_i, \sin \theta_i)$ a unit polarization vector that randomly changes direction over time:
    \begin{equation}
        \partial_t \theta_i = \eta_i(t), \quad \langle \eta_i(t) \; \eta_j(t') \rangle = 2 D_r \delta_{ij}\delta(t-t')
    \end{equation}
    
    The cell centers are displaced using a forward Euler discretization of eq.~\eqref{eq_motion_cellcenter} and the Voronoi tessellation is computed with the new position of the cell centers.

    \subsection{Coupling between mechanics and growth}
    While we can infer general trends for the effect of mechanics on growth such as pressure-induced inhibition of growth \cite{montel_stress_2011} or regulation of homeostatic density \cite{eisenhoffer_crowding_2012}, pinpointing the exact influence of mechanical variables on individual cells proves challenging.
    Experimentally, researchers have access to macroscopic measures such as the force on a cell assembly or the force exerted on the substrate using traction force microscopy. However, quantifying bulk properties such as the stress within a cell monolayer remains elusive.
    The stress within a monolayer can be computed using imaging-based inference methods \cite{ishihara_bayesian_2012,yang_correlating_2017}, but since the methods rely on formulating models similar to the one we are using, it does not provide new information. 

    In the absence of a clear coupling between mechanics and growth dictated by experiments, we can think of a generalized linear response of growth to mechanical changes. 
    If one thinks of the mechanical parameters $a_0,p_0$ in the energy functional as some kind of target homeostatic value for the tissue, then the birth and death rates of a cell $r_i$ can be expressed as
    \begin{align}
        r^b_i &= {r^b_i}^0 + \lambda \max\left(0, a-a_0 + \beta (p-p_0) \right) \label{eq:birthrate-general}\\
        r^d_i &= {r^d_i}^0 - \lambda \min \left(0, a-a_0 + \beta (p-p_0) \right) \label{eq:deathhrate-general}
    \end{align}
    The first term corresponds to the basal components of the birth and death rates, respectively ${r^b_i}^0$ and ${r^d_i}^0$, which do not depend on mechanical parameters.
    The second term is a linear coupling between the mechanical state of the system and the growth rate $r^b_i-r^d_i$, with a coupling coefficient $\lambda$ and a parameter $\beta$ that weighs the contribution of the perimeter with respect to the area.
    The $\min$ and $\max$ in the death and birth rates ensure that each rate is positive.
    Higher coupling terms are neglected.

    In this first work, we focus on the case $\beta=0$, which means that the growth rate depends only on the area of the cells, or equivalently on the density in the tissue.
    This particular coupling is relevant in epithelial monolayers, where there is more extrusion of cells in denser regions \cite{eisenhoffer_crowding_2012}.
    This coupling is related to the hydrostatic pressure defined as $\at{\partial E / \partial A_i}{P_i} = -K(A_i-A_i^0)$.
    The hydrostatic pressure represents the normal forces exerted on the edges of the cells \cite{yang_correlating_2017}.

    For simplicity, let us consider the case with the same basal birth and death rates for all cells, ${r^b_i}^0={r^d_i}^0 = r^0$.
    In that case, the rates are given by:
    \begin{align}
        r^b_i &= r^0 + \lambda \max\left(0,a_i-a_0\right) \label{eq:birthrate-area}\\
        r^d_i &= r^0 - \lambda \min\left(0, a_i-a_0\right) \label{eq:deathhrate-area}
    \end{align}
    With our normalization scheme, the average growth rate at time $t$ is fixed by eq.~\eqref{eq:volumeConservation} and depends on the number of cells:
    \begin{equation}
       \langle g_i \rangle =   \langle r^b_i - r^d_i\rangle = \lambda \left(\langle a_i \rangle -a_0 \right) = \lambda \left(\frac{N(t=0)}{N(t)}-a_0 \right)
    \end{equation}
    The steady state is reached when the average growth rate $\langle g_i \rangle$ vanishes and the steady-state cell number is set by the initial cell number and the preferred area $N_{SS} = N(t=0)/a_0$.
    We represent the evolution of cell number over time using different growth couplings in \cref{fig2}.

    Another mechanical observable associated with growth is the pressure \cite{montel_stress_2011,delarue_compressive_2014,basan_homeostatic_2009}, which corresponds to a special value of $\beta$.
    The pressure is defined as the opposite of the isotropic part of the stress tensor, which can be determined from the forces using the virtual work principle \cite{nestor-bergmann_relating_2018}. 
    The dimensionless pressure $\mathcal{P}$, normalized by $KL^2/N_0$, is given by:
    \begin{equation}
        \mathcal{P} = -(a-a^0) - \frac{\Gamma p}{2a}(p-p^0).
        \label{eq_totalpressure}
    \end{equation}
    To linear order in $a-a^0$ and $p-p^0$, it corresponds to $\beta = \Gamma p^0/(2 a^0)$.
    The birth and death rates of cell $i$ are expressed as:
    \begin{align}
        r^b_i &= r^0 + \lambda \max\left(0,\mathcal{P}_i-\mathcal{P}_i^h\right) \label{eq:birthrate-pressure}\\
        r^d_i &= r^0 - \lambda \min\left(0, \mathcal{P}_i-\mathcal{P}_i^h \right) \label{eq:deathhrate-pressure}
    \end{align}
    where $\mathcal{P}^h$ is viewed as the homeostatic pressure, since it determines the value of the pressure in steady state. To keep the number of initial cells constant, there is a unique value of $\mathcal{P}^h$ that depends on the mechanical parameters.

    In a future work, a comparison between different couplings between the growth rate and mechanical properties and their effect on cell competition will be studied.
    Notably, the limit $\beta \to \infty$ is unstable. When $\beta \gg 1$, a larger $p$ leads to a higher growth rate, regardless of the area. Since there is no constraint on the perimeter, this leads to infinitely elongated cells.

\section{Results}
    \subsection{Growth of an homogeneous tissue}
\begin{figure}
        \centering
        \includegraphics[width=\textwidth]{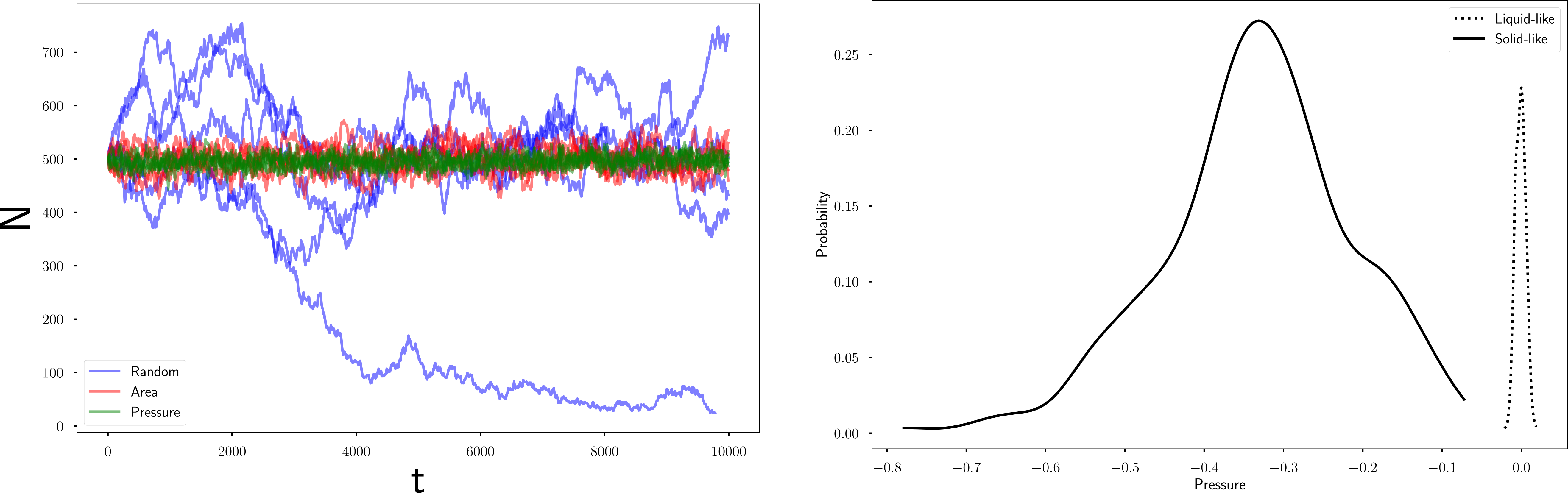}
        \caption{Left panel: evolution of the number of cells over time for different regulation of the birth-death process. We notice that purely stochastic process (blue) leads to large deviations of the cell number over time, while birth-death processes regulated by mechanics, either through the density (red) or the pressure (green), lead to a steady-state value of the number of cells.
        Right panel: distribution of the time-averaged pressure in the tissue. We notice that in the solid-like state (solid line) there is a negative pre-stress value that is not relaxed. However, the liquid-like state (dashed line) is able to relax this pre-stress.}
        \label{fig2}
    \end{figure}

    The cell number over time from different couplings between mechanics and birth-death rates is shown in \cref{fig2}. 
    In all cases, ${r^b_i}^0={r^d_i}^0$ for all cells. 
    We looked at three cases, $\lambda=0$ (no coupling to mechanics), $\beta = 0$ which is density regulation, and coupling to pressure. 
    In each case, there are five realizations of the stochastic birth-death process using a Gillespie algorithm.
    We see that when there is no coupling to mechanics, the stochastic birth-death process does not reach a steady state. The cell number is only subject to demographic noise and there is no effective carrying capacity.
    However, coupling to mechanical forces stabilizes the number of cells to a value that depends on the mechanical parameters. 
    We therefore argue that a mechanical feedback is necessary to maintain a fixed average cell number when the birth-death process is noisy.
        
    \subsection{Invasion by a mutant population} \label{sec:results_area}
    After rescaling physical units, our model contains a total of 15 parameters: 7 energy parameters, 4 growth rate parameters, and 4 motility parameters. 
    Since we are interested in the competition between mutants and residents, we decided to fix some parameters of the resident population.
    We end up with a total of 8 parameters involved in the competition.
    The parameters and the effect of their change are recapitulated in \cref{tab:parameters}.

\begin{table}[htp]
    \centering
    \begin{adjustbox}{width=\textwidth}
    \begin{tabular}{p{.12\textwidth}|p{.44\textwidth}|p{.44\textwidth}}
        Parameter & Description & Effect on fitness of mutant \\
         \hline
        $a_{r,m}^0$ & Preferred area & Fitness advantage for mutants when $a_m^0 < a_r^0$. Resident value fixed to $a_{r}^0=1$\\ \hline
        $p_{r,m}^0$ & Preferred perimeter & Sets the phase of the system.  Fitness advantage if $p_r^0>p_m^0$ \\ \hline
        $\Gamma_{r,m}$ & Perimeter elasticity & No apparent effect. Resident value fixed to $\Gamma_{r}=1$ \\ \hline
        $K_m$ & Area elasticity of mutant & No apparent effect \\ \hline
        $\lambda_{r,m}$ & Strength of mechanical coupling to growth & No apparent effect. Resident value fixed to $\lambda_r =1$ \\ \hline
        $r^0_{r,m}$ & Intrinsic growth rate & Provides a fitness advantage when $r_m^0 > r^0_r$. Resident value fixed to $r^0_{r}=0$ \\ \hline
        $v_{r,m}^0$ & Self-advection & No effect. Resident value fixed to $v_{r}^0 = 0.05$ \\ \hline
        $D_{r,m}$ & Rotational diffusion (inverse persistence time) & No effect. Resident value fixed to $D_r =1$
    \end{tabular}
    \end{adjustbox}
    \caption{Simulation parameters and their effect on fitness}
    \label{tab:parameters}
\end{table}

\begin{figure}[h]
    \centering
    \includegraphics[width=1.2\textwidth,center]{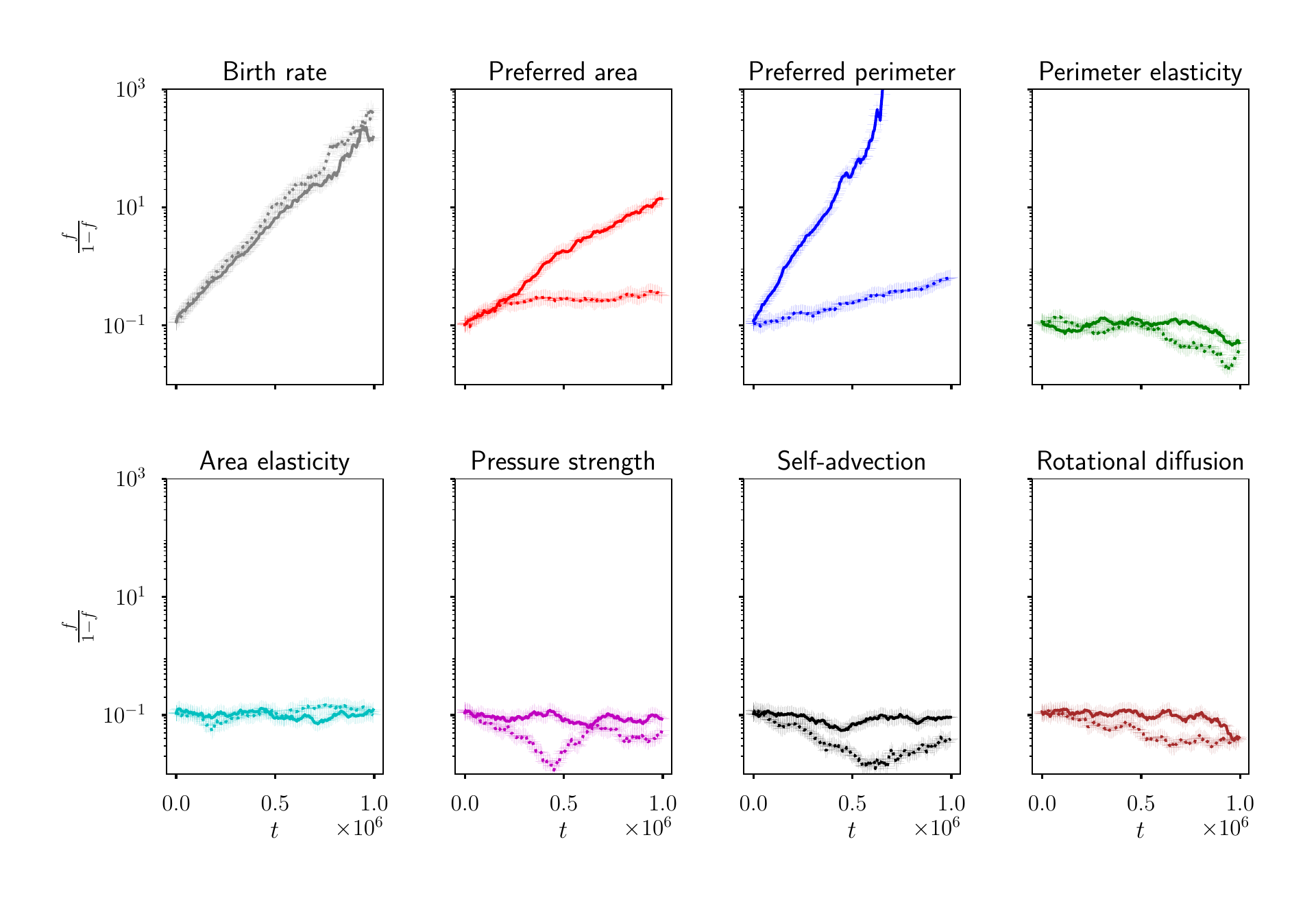}
    \caption{Evolution of the fraction of mutants from an initial droplet of fraction $f=0.1$, when the birth and death rates are controlled by the area, as stated in \cref{eq:birthrate-area,eq:deathhrate-area}. Each panel corresponds to an average of 10 identical simulations, where there is a small change in one parameter for the mutant population. Solid lines are for simulations where the resident is in a solid-like state with $p_r^0 = 3.3$, and dotted lines are for simulations where the resident is in a liquid-like state with $p_r^0 =4$}
    \label{fig:invasion}
\end{figure}

The parameters of resident and mutant cells can be thought of as phenotypic traits that can be changed by mutations.
We are interested in the evolutionary outcome of a change in any of these phenotypic traits. 
To test this, we performed numerical invasion experiments in which we introduce a small fraction $f$ of mutants among a resident population. 
The mutants have a small change of the order of $10\%$ in one of the parameters compared to the residents, and we performed 8 invasion experiments, one for each of the competition parameters (see \cref{tab:parameters}).
Looking at the evolution of the fraction $f$ of mutants over time, we can determine if the change is advantageous -- $f$ increases on average, eventually leading to fixation -- or deleterious -- $f$ decreases on average leading to extinction.
The evolution of the ratio between the mutant fraction and the resident fraction is represented in \cref{fig:invasion} on a semi-log scale.
On average, the mutant fraction evolves according to a logistic equation $\partial_t f = s f (1-f)$,
where $s = g_m - g_r$ is the growth rate difference between mutants and residents.
If $s$ does not depend on the mutant fraction, the solution to the logistic equation is simply $\ln (f/(1-f)) = s t$. 
Therefore, the straight lines on figure~\ref{fig:invasion} correspond to growth rate differences that do not depend on the mutant fraction.
From figure~\ref{fig:invasion} we identify 3 parameters that provide a fitness difference, while the other 5 parameters do not significantly change the fraction of mutant.
Interestingly, for the 3 parameters that provide a change in fitness, this change is constant over time, since $\ln(f/(1-f))$ is a linear function of time.

The effect of the basal rate of the mutant $r_m^0$ is trivial since it provides a constant growth rate difference.
For the two geometric parameters, the preferred area $a^0$ and the preferred perimeter $p^0$, it could be expected that the growth rate difference generated would change as the mutant fraction evolves. 
However, we find that the growth rates of the resident and mutant population change over time, but the difference between the two remain constant, mainly due to the volume constraint. 
Let us take for example the case where the preferred area is varied, and the mutant has a smaller preferred area $a_m^0 = 0.9$ while the resident preferred area is $a_r^0 = 1$.
The volume constraint fixes the average area of the entire population by the number of cells, according to eq.~\eqref{eq:volumeConservation}.
Initially, the average area of each cell is fixed to unity. 
The mutant cells introduced are trying to be smaller than the resident cells to minimize their energy.
With the average area fixed roughly to unity, the mutant cells will be greater than their preferred area, but so will the resident cells that grow in size to compensate for the smaller mutant cells. 
This effect increases the growth rate of both mutant and resident cells, depending on their relative fraction and the total number of cells.
However, the difference between the two growth rates does not depend on the fraction nor the total number of cells.
A more detailed derivation from energetic arguments is provided in section \ref{sec:theory}.
The argument for a change in the preferred perimeter $p^0$ follows the same path, since it effectively changes the realized area of the cells.
We find that the effect of the geometric parameters $a^0$ and $p^0$ on fitness is stronger in the solid-like phase than it is in the liquid-like phase.
We attribute this effect to the broader distribution of cell areas in the solid-like phase with respect to the liquid-like phase (see \cref{fig:snapshot}).

A small change in the strength of the coupling between the area and the growth rate $\lambda$ seems to have no effect on the evolution of the population. 

Mechanical elasticities $\Gamma,K$ play little role in the fitness when they are varied alone. However, we show in section~\ref{sec:theory} that they control the magnitude of growth rate difference, even though they cannot create such a difference. 

\subsection{Evolutionary outcome determined by a non-growing mixture} \label{sec:theory}
Since the rate of change of the mutant fraction is constant over time, we can determine the fate of the competition between mutant and resident strains by simulating a mixture of the two strains at a constant fraction, without growth.
We simulate a 50-50 mixture of mutant and resident strains without growth and compute the average realized areas for mutant and resident strains, resp. $\langle a_m \rangle$ and $\langle a_r \rangle$, which gives the average growth rate difference between mutant and resident strains $s = \lambda (\langle a_m \rangle - a_m^0 - (\langle a_r \rangle - a_r^0))$. 
The results are shown in \cref{fig:estimation}. The sign of the growth rate difference is well predicted by simulating a mixture without growth, but the actual value of the growth rate is somewhat different. This discrepancy could be explained by the fluctuations in cell number induced by the fluctuations in the birth-death process, since a change in cell number changes the average value of the area (see \cref{eq:volumeConservation}).
\begin{figure}
    \centering
    \includegraphics[width=\textwidth]{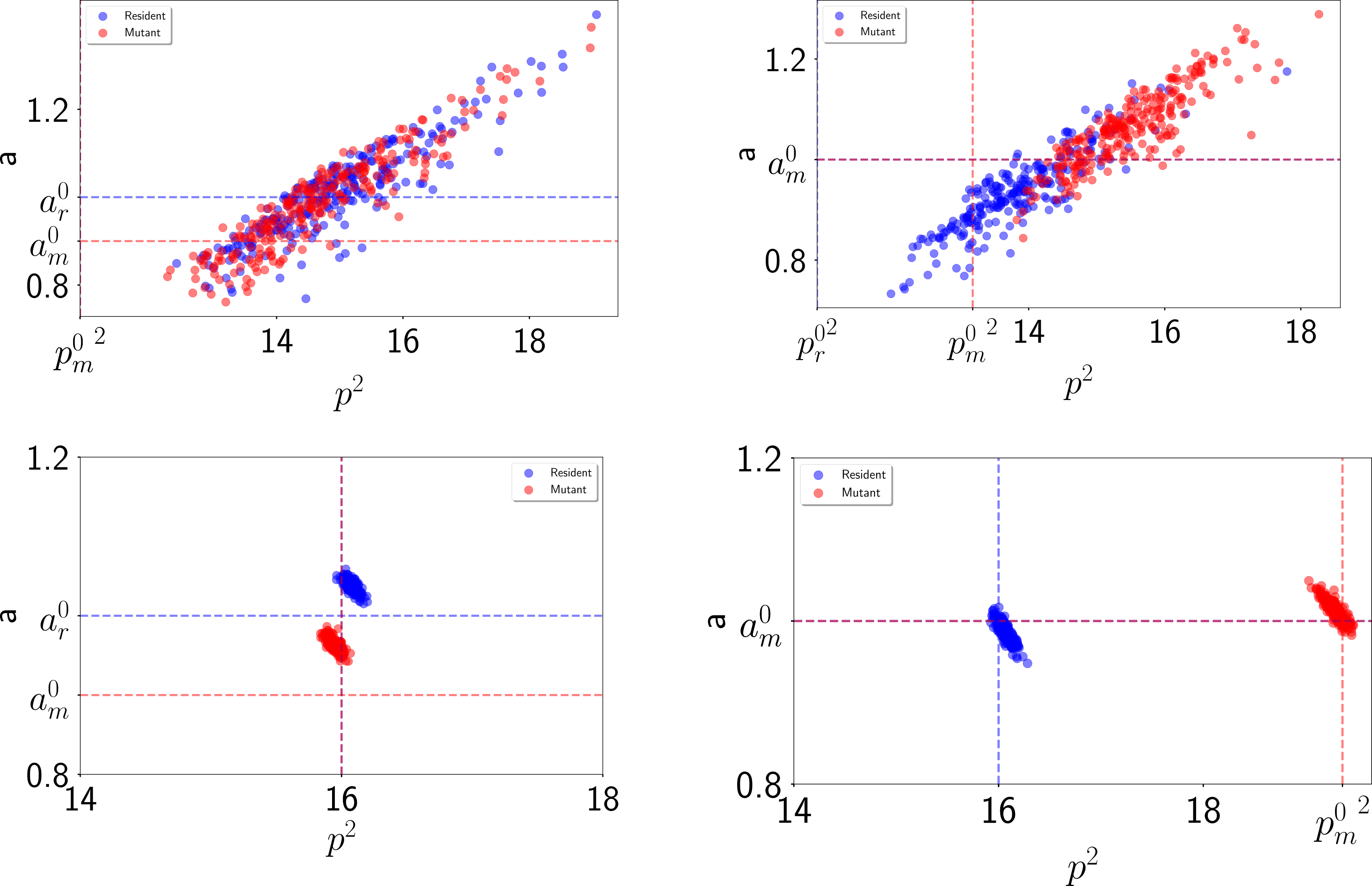}
    \caption{Distribution of areas and perimeter for solid (upper row) and liquid (lower row) 50-50 mixtures of resident and mutants averaged over time. In the left column, mutants have a preferred area of $a_m^0 = 0.9$ while the one of residents is $a_r^0 =1$ (horizontal dashed lines). All other parameters are equal. In the right column, mutants have a preferred perimeter of $p_m^0 = 0.9$ while residents have a preferred perimeter $p_r^0 =1$ (vertical dashed lines). All other parameters are equal and have their reference value.
    }
    \label{fig:snapshot}
\end{figure}
\begin{figure}
    \centering
    \includegraphics[width=.6\textwidth]{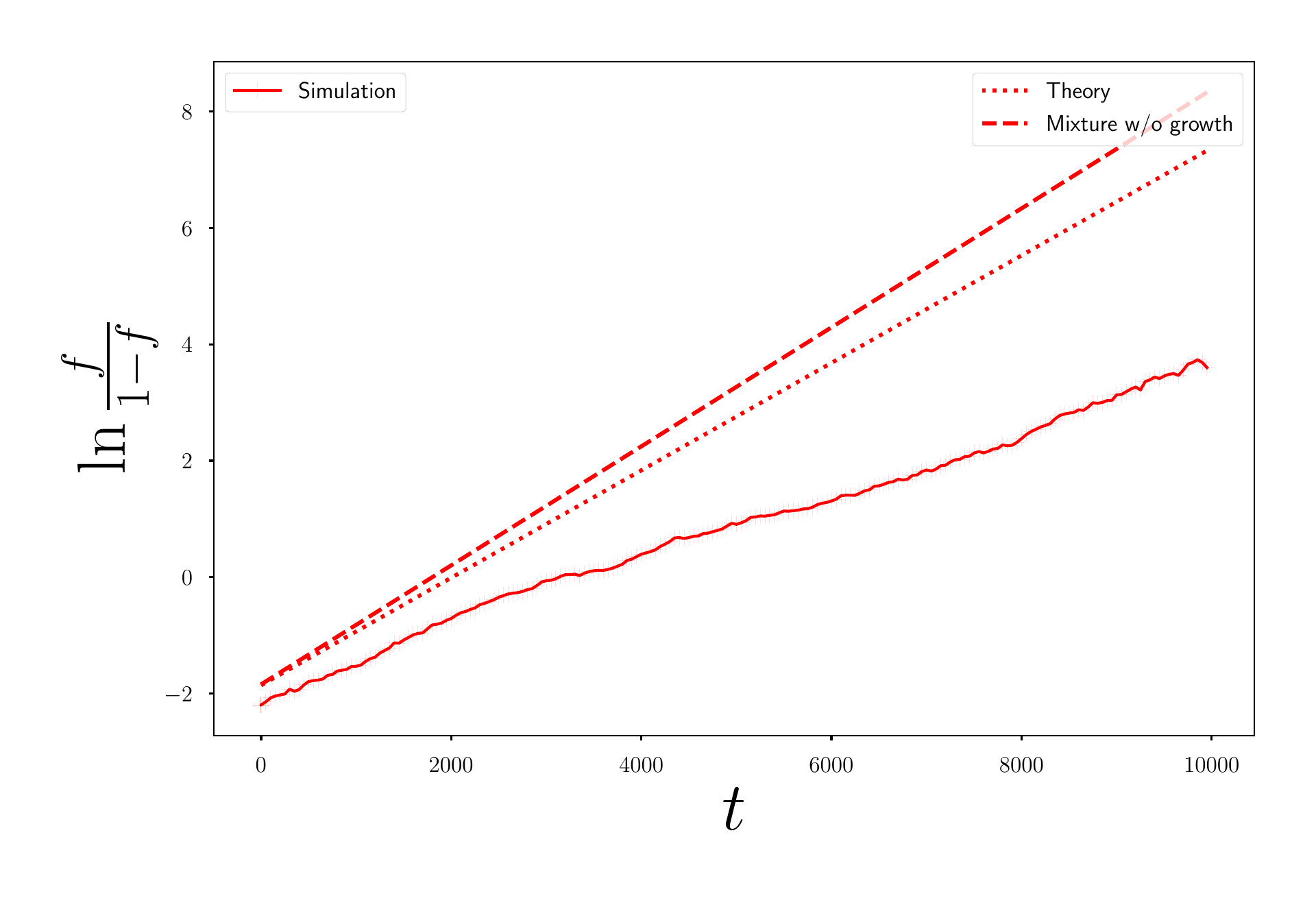}
    \caption{Evolution of the fraction of mutants over time when the preferred area of the mutant is $a_m^0 = 0.9$ while the one of residents is $a_r^0 =1$. The solid line represents the simulation of the birth-and-death process in a solid state $p_r^0 = p_m^0 = 3.63$. All other parameters are equal and have their reference value.
    The dotted line is the estimation from the mean-field approximation given by \cref{eq:s_meanfield_area} and the dashed line is an estimation using the area difference given by a mixture without growth.
    We see that although both estimation predict the same evolutionary outcome, they overestimate the the rate of evolution.}
    \label{fig:estimation}
\end{figure}
Experimentally, this mean the evolutionary information can be obtained at any point in time from an heterogeneous population of cells, by averaging the distribution of areas.

In this section, we present an energy argument to explain why the growth rate difference between mutants and residents does not depend on the fraction. 
Consider a mixture with residents having a preferred area $a_r^0 = 1$ and mutants having a preferred area $a_m^0$. Let us assume that both residents and mutants have the same preferred perimeter $p^0$. The initial population is composed of $N_0$ cells in a fixed volume $L^2$ (an average cell initially has an area of value 1).
We consider that at time $t$ there are $N_t$ cells, a fraction $f = N_m/N_t$ of mutant cells, and $1-f = N_r/N_t$ resident cells. 
Consider a mean-field approximation to solve the evolution of identical average cells.
In that case, the volume conservation at time $t$ reads:
\begin{equation}
    (1-f)a_r + f a_m = \frac{N_0}{N_t}
    \label{eq:volum_conserv_area}
\end{equation} 
Equation~\eqref{eq:volum_conserv_area} sets the average cell size to be $N_0/N_t$. However, this value does not minimize the energy of the mixture with two different preferred areas.
We assume that there are small deviations $\delta a_m, \delta a_r \ll \frac{N_0}{N_t}$ w.r.t the average value, and write the areas as $a_m = N_0/N_t + \delta a_m $ and $a_r = N_0/N_t + \delta a_r $.
If the deviations in areas are small, to first order the perimeters are
\begin{align}
    p_m \approx p^0 + c\delta a_m \\
    p_r \approx p^0 + c\delta a_r 
\end{align}
We can make the assumption that the proportionality coefficient $c$ is the same for both mutants and residents since having the same preferred perimeter means they are similar shaped polygons.

With this perturbation, there is only one degree of freedom, and we can choose it to be $\delta a_m$.
By minimizing the energy with respect to $\delta a_m$, we obtain the mean-field equilibrium areas for the mutant and resident cells.
From these areas, we determine the mean-field prediction for the fitness advantage when the preferred area is varied:
\begin{equation}
    s_{MF}^A =  -\lambda \Delta a^0\frac{\Gamma c^2 }{1+\Gamma c^2 },
    \label{eq:s_meanfield_area}
\end{equation}
with $\Delta a^0 = a_m^0-a_r^0$ is the difference in preferred areas. 
The sign of $\Delta a^0$ determines the evolutionary outcome and supports the finding from the invasion simulation that $a_m^0<a_r^0$ leads to a fitness advantage.

The exact same thing can be done when the preferred perimeter is varied, and gives
\begin{equation}
    s_{MF}^P =  \lambda \Delta p^0\frac{\Gamma c }{1+\Gamma c^2},
    \label{eq:s_meanfield_p0}
\end{equation}
with $\Delta p^0 = p_m^0-p_r^0$ is the difference in preferred perimeters. Similarly, the sign of $\Delta p^0$ determines the evolutionary outcome and supports the finding from the invasion simulation that $p_m^0>p_r^0$ leads to a fitness advantage.

Both \cref{eq:s_meanfield_area,eq:s_meanfield_p0} confirm that the fitness is independent of frequency. The growth rate of each type depends on the frequency of the mutant, but the difference between the growth rate of each does not depend on the frequency. 

We can also note that although the normalized perimeter elasticity $\Gamma$ does not play a role without a mutation in the preferred area or perimeter, it controls the rate of evolution when such a mutation exists.
The greater $\Gamma$ is, the faster evolution occurs. A hand-wavy argument is that a high perimeter elasticity fixes the shape of the cells, making it harder to deform in the network to accommodate changes in area. 

Both rates depend on the coefficient $c$ that relates a change in area to a change in perimeter.
For a regular n-sided polygon, there is a relationship $a = 1/c_n p^2$, with $c_n = 4n\tan(\pi/n)$.
In the solid-like phase, when averaging over time the distribution of areas and perimeters, we indeed observe a linear relationship between the area and the squared perimeter, as seen in the first row of \cref{fig:snapshot}.
Using the coefficient for regular hexagons $c_6 \approx 13.9$ fits well the distribution of areas and perimeters in the solid phase.
In the liquid-like phase, the shapes of the cells are not well described by regular polygons, and there is no a priori relationship between the area and the perimeter.
The simulation of a mixture of resident and mutant cells shows that there is an area change induced by a change in preferred perimeter and vice versa.
This change in area is smaller than for the solid-like phase, which explains the smaller rate of change of the mutant fraction in the liquid-like phase compared to the solid-like phase observed in \cref{fig:invasion}.

\section{Discussion}
Mechanics play a role in the regulation of growth in tissue \cite{eisenhoffer_crowding_2012,montel_stress_2011,delarue_compressive_2014}, and thus can be a driver of cell competition, for example in the case of cancer invasion \cite{levayer_solid_2020}.
Using an agent-based model of tissue mechanics with mechanically-regulated growth bridges the gap between evolutionary models of tumors \cite{bozic_evolutionary_2013,waclaw_spatial_2015} that are concerned with the genetic diversity in tumors and macroscopic hydrodynamic or elastic models that look at deformations and growth induced by heterogeneous cells \cite{taloni_role_2015,ben_amar_contour_2011}.

Mechanically-regulated growth in a homogeneous tissue controls the average number of cells in the system even in the presence of demographic noise, by effectively providing a carrying capacity.
To study competition in heterogeneous tissues, we simulated invasion scenarios by different mechanical mutants and have identified that the geometrical parameters in the Voronoi model, the preferred area $a_0$ and the preferred perimeter $p_0$, confer a fitness advantage to the mutants, while other parameters do not seem to play a role. 
A change in preferred area can be interpreted macroscopically as a change of homeostatic pressure, where lowering the preferred area is similar to increasing the homeostatic pressure, i.e. the pressure at which the tissue is in a homeostatic growth state. 
In this light, this model agrees with hydrodynamic models of tumor growth coupled to mechanics \cite{basan_homeostatic_2009,shraiman_mechanical_2005}.
The fitness advantage, when it exists, is constant over time, which means that it does not depend on the frequency of the mutant in the tissue. Therefore, we can explain the long-term evolutionary fate of a mutant -- fixation or extinction -- either by energy consideration on a tissue made of average cells or by simulating a mixture of mutants and residents without growth.
This is relevant for experiments since it means that one does not have to conduct a full evolutionary experiment to infer the outcome, it is sufficient to know the average state of tissue at a certain time point.

In this work, we studied in detail the case in which the growth rate depends on the cell area, and we should note that the results obtained depend on the chosen mechanical coupling.
For example, if one were to consider the birth rate of a cell to be proportional only to the realized area of the cell instead of the difference between the realized and preferred areas, a change in preferred area would have opposite effects on the mutant fraction, as a larger preferred area would just lead to larger cells. 
A systematic study in future works of possible couplings between mechanics and growth could help discriminate which coupling is most relevant to experimental realizations. 
We found that one of the key features of the SPV model, the jamming-unjamming transition~\cite{bi_density-independent_2015,bi_motility-driven_2016,chiang_glass_2016}, is not a key driver of cell competition.
We believe that there are two reasons why there are no fundamental differences between the two phases. 
The first is that the growth process essentially fluidizes the tissue \cite{ranft_fluidization_2010} and gives rise to a non-vanishing effective diffusion, effectively unjamming the jammed phase. 
The second reason is that over the time of the simulation the system is essential well-mixed, there is no sustained spatial difference between the mutants and the residents. 

A next step for this work would be to study the effect of mechanics on genetic diversity. At larger spatial scales, one would observe nontrivial growth-induced pressure gradients and traveling fronts, which could change important aspects of the competition \cite{ranft_mechanically_2014, korolev_genetic_2010, birzu_fluctuations_2018}.

\begin{acknowledgments}
We thank Daniel Sussman for developing cellGPU and making it easily accessible to the scientific community, and the members of the Theoretical Biophysics group at BU for helpful discussions and feedback. LB and KK were supported by NIGMS Grant No. 1R01GM138530-01
\end{acknowledgments}

\bibliographystyle{ieeetr}
\bibliography{Library}

\begin{thebibliography}{10}

\bibitem{tamori_cell_2011}
Y.~Tamori and W.-M. Deng, ``Cell competition and its implications for development and cancer,'' {\em Journal of Genetics and Genomics}, vol.~38, pp.~483--495, Oct. 2011.

\bibitem{morata_minutes_1975}
G.~Morata and P.~Ripoll, ``Minutes: {Mutants} of {Drosophila} autonomously affecting cell division rate,'' {\em Developmental Biology}, vol.~42, no.~2, pp.~211--221, 1975.

\bibitem{korolev_turning_2014}
K.~S. Korolev, J.~B. Xavier, and J.~Gore, ``Turning ecology and evolution against cancer,'' {\em Nature Reviews Cancer}, vol.~14, pp.~371--380, May 2014.

\bibitem{sender_distribution_2021}
R.~Sender and R.~Milo, ``The distribution of cellular turnover in the human body,'' {\em Nature Medicine}, vol.~27, pp.~45--48, Jan. 2021.

\bibitem{guillot_mechanics_2013}
C.~Guillot and T.~Lecuit, ``Mechanics of {Epithelial} {Tissue} {Homeostasis} and {Morphogenesis},'' {\em Science}, vol.~340, pp.~1185--1189, June 2013.

\bibitem{matamoro-vidal_multiple_2019}
A.~Matamoro-Vidal and R.~Levayer, ``Multiple {Influences} of {Mechanical} {Forces} on {Cell} {Competition},'' {\em Current Biology}, vol.~29, pp.~R762--R774, Aug. 2019.

\bibitem{levayer_solid_2020}
R.~Levayer, ``Solid stress, competition for space and cancer: {The} opposing roles of mechanical cell competition in tumour initiation and growth,'' {\em Seminars in Cancer Biology}, vol.~63, pp.~69--80, June 2020.

\bibitem{wagstaff_mechanical_2016}
L.~Wagstaff, M.~Goschorska, K.~Kozyrska, G.~Duclos, I.~Kucinski, A.~Chessel, L.~Hampton-O’Neil, C.~R. Bradshaw, G.~E. Allen, E.~L. Rawlins, P.~Silberzan, R.~E. Carazo~Salas, and E.~Piddini, ``Mechanical cell competition kills cells via induction of lethal p53 levels,'' {\em Nature Communications}, vol.~7, p.~11373, Apr. 2016.

\bibitem{aragona_mechanical_2013}
M.~Aragona, T.~Panciera, A.~Manfrin, S.~Giulitti, F.~Michielin, N.~Elvassore, S.~Dupont, and S.~Piccolo, ``A {Mechanical} {Checkpoint} {Controls} {Multicellular} {Growth} through {YAP}/{TAZ} {Regulation} by {Actin}-{Processing} {Factors},'' {\em Cell}, vol.~154, pp.~1047--1059, Aug. 2013.

\bibitem{balasubramaniam_investigating_2021}
L.~Balasubramaniam, A.~Doostmohammadi, T.~B. Saw, G.~H. N.~S. Narayana, R.~Mueller, T.~Dang, M.~Thomas, S.~Gupta, S.~Sonam, A.~S. Yap, Y.~Toyama, R.-M. Mège, J.~M. Yeomans, and B.~Ladoux, ``Investigating the nature of active forces in tissues reveals how contractile cells can form extensile monolayers,'' {\em Nature Materials}, vol.~20, pp.~1156--1166, Aug. 2021.

\bibitem{montel_stress_2011}
F.~Montel, M.~Delarue, J.~Elgeti, L.~Malaquin, M.~Basan, T.~Risler, B.~Cabane, D.~Vignjevic, J.~Prost, G.~Cappello, and J.-F. Joanny, ``Stress {Clamp} {Experiments} on {Multicellular} {Tumor} {Spheroids},'' {\em Physical Review Letters}, vol.~107, p.~188102, Oct. 2011.

\bibitem{eisenhoffer_crowding_2012}
G.~T. Eisenhoffer, P.~D. Loftus, M.~Yoshigi, H.~Otsuna, C.-B. Chien, P.~A. Morcos, and J.~Rosenblatt, ``Crowding induces live cell extrusion to maintain homeostatic cell numbers in epithelia,'' {\em Nature}, vol.~484, pp.~546--549, Apr. 2012.

\bibitem{shraiman_mechanical_2005}
B.~I. Shraiman, ``Mechanical feedback as a possible regulator of tissue growth,'' {\em Proceedings of the National Academy of Sciences}, vol.~102, pp.~3318--3323, Mar. 2005.

\bibitem{basan_homeostatic_2009}
M.~Basan, T.~Risler, J.~Joanny, X.~Sastre‐Garau, and J.~Prost, ``Homeostatic competition drives tumor growth and metastasis nucleation,'' {\em HFSP Journal}, vol.~3, pp.~265--272, Aug. 2009.

\bibitem{ranft_mechanically_2014}
J.~Ranft, M.~Aliee, J.~Prost, F.~Jülicher, and J.-F. Joanny, ``Mechanically driven interface propagation in biological tissues,'' {\em New Journal of Physics}, vol.~16, p.~035002, Mar. 2014.

\bibitem{farhadifar_influence_2007}
R.~Farhadifar, J.-C. Röper, B.~Aigouy, S.~Eaton, and F.~Jülicher, ``The {Influence} of {Cell} {Mechanics}, {Cell}-{Cell} {Interactions}, and {Proliferation} on {Epithelial} {Packing},'' {\em Current Biology}, vol.~17, no.~24, pp.~2095--2104, 2007.

\bibitem{alt_vertex_2017}
S.~Alt, P.~Ganguly, and G.~Salbreux, ``Vertex models: from cell mechanics to tissue morphogenesis,'' {\em Philosophical Transactions of the Royal Society B: Biological Sciences}, vol.~372, p.~20150520, May 2017.

\bibitem{bi_density-independent_2015}
D.~Bi, J.~H. Lopez, J.~M. Schwarz, and M.~L. Manning, ``A density-independent rigidity transition in biological tissues,'' {\em Nature Physics}, vol.~11, pp.~1074--1079, Dec. 2015.

\bibitem{bi_motility-driven_2016}
D.~Bi, X.~Yang, M.~C. Marchetti, and M.~L. Manning, ``Motility-{Driven} {Glass} and {Jamming} {Transitions} in {Biological} {Tissues},'' {\em Physical Review X}, vol.~6, p.~021011, Apr. 2016.

\bibitem{chiang_glass_2016}
M.~Chiang and D.~Marenduzzo, ``Glass transitions in the cellular {Potts} model,'' {\em EPL (Europhysics Letters)}, vol.~116, p.~28009, Oct. 2016.

\bibitem{mitchel_primary_2020}
J.~A. Mitchel, A.~Das, M.~J. O’Sullivan, I.~T. Stancil, S.~J. DeCamp, S.~Koehler, O.~H. Ocaña, J.~P. Butler, J.~J. Fredberg, M.~A. Nieto, D.~Bi, and J.-A. Park, ``In primary airway epithelial cells, the unjamming transition is distinct from the epithelial-to-mesenchymal transition,'' {\em Nature Communications}, vol.~11, p.~5053, Dec. 2020.

\bibitem{park_unjamming_2015}
J.-A. Park, J.~H. Kim, D.~Bi, J.~A. Mitchel, N.~T. Qazvini, K.~Tantisira, C.~Y. Park, M.~McGill, S.-H. Kim, B.~Gweon, J.~Notbohm, R.~Steward~Jr, S.~Burger, S.~H. Randell, A.~T. Kho, D.~T. Tambe, C.~Hardin, S.~A. Shore, E.~Israel, D.~A. Weitz, D.~J. Tschumperlin, E.~Henske, S.~T. Weiss, M.~L. Manning, J.~P. Butler, J.~M. Drazen, and J.~J. Fredberg, ``Unjamming and cell shape in the asthmatic airway epithelium,'' {\em Nature Materials}, vol.~14, pp.~1040--1048, Oct. 2015.

\bibitem{atia_are_2021}
L.~Atia, J.~J. Fredberg, N.~S. Gov, and A.~F. Pegoraro, ``Are cell jamming and unjamming essential in tissue development?,'' {\em Cells \& Development}, vol.~168, p.~203727, Dec. 2021.

\bibitem{graner_simulation_1992}
F.~Graner and J.~A. Glazier, ``Simulation of biological cell sorting using a two-dimensional extended {Potts} model,'' {\em Physical Review Letters}, vol.~69, pp.~2013--2016, Sept. 1992.

\bibitem{sussman_anomalous_2018}
D.~M. Sussman, M.~Paoluzzi, M.~Cristina~Marchetti, and M.~Lisa~Manning, ``Anomalous glassy dynamics in simple models of dense biological tissue,'' {\em EPL (Europhysics Letters)}, vol.~121, p.~36001, Feb. 2018.

\bibitem{sussman_no_2018}
D.~M. Sussman and M.~Merkel, ``No unjamming transition in a {Voronoi} model of biological tissue,'' {\em Soft Matter}, vol.~14, no.~17, pp.~3397--3403, 2018.

\bibitem{sussman_cellgpu_2017}
D.~M. Sussman, ``{cellGPU}: massively parallel simulations of dynamic vertex models,'' {\em Computer Physics Communications}, vol.~219, pp.~400--406, Oct. 2017.
\newblock arXiv: 1702.02939.

\bibitem{ishihara_bayesian_2012}
S.~Ishihara and K.~Sugimura, ``Bayesian inference of force dynamics during morphogenesis,'' {\em Journal of Theoretical Biology}, vol.~313, pp.~201--211, Nov. 2012.

\bibitem{yang_correlating_2017}
X.~Yang, D.~Bi, M.~Czajkowski, M.~Merkel, M.~L. Manning, and M.~C. Marchetti, ``Correlating cell shape and cellular stress in motile confluent tissues,'' {\em Proceedings of the National Academy of Sciences}, vol.~114, pp.~12663--12668, Nov. 2017.

\bibitem{delarue_compressive_2014}
M.~Delarue, F.~Montel, D.~Vignjevic, J.~Prost, J.-F. Joanny, and G.~Cappello, ``Compressive {Stress} {Inhibits} {Proliferation} in {Tumor} {Spheroids} through a {Volume} {Limitation},'' {\em Biophysical Journal}, vol.~107, pp.~1821--1828, Oct. 2014.

\bibitem{nestor-bergmann_relating_2018}
A.~Nestor-Bergmann, G.~Goddard, S.~Woolner, and O.~E. Jensen, ``Relating cell shape and mechanical stress in a spatially disordered epithelium using a vertex-based model,'' {\em Mathematical Medicine and Biology: A Journal of the IMA}, vol.~35, pp.~i1--i27, Apr. 2018.

\bibitem{bozic_evolutionary_2013}
I.~Bozic, J.~G. Reiter, B.~Allen, T.~Antal, K.~Chatterjee, P.~Shah, Y.~S. Moon, A.~Yaqubie, N.~Kelly, D.~T. Le, E.~J. Lipson, P.~B. Chapman, L.~A. Diaz, B.~Vogelstein, and M.~A. Nowak, ``Evolutionary dynamics of cancer in response to targeted combination therapy,'' {\em eLife}, vol.~2, p.~e00747, June 2013.

\bibitem{waclaw_spatial_2015}
B.~Waclaw, I.~Bozic, M.~E. Pittman, R.~H. Hruban, B.~Vogelstein, and M.~A. Nowak, ``A spatial model predicts that dispersal and cell turnover limit intratumour heterogeneity,'' {\em Nature}, vol.~525, pp.~261--264, Sept. 2015.

\bibitem{taloni_role_2015}
A.~Taloni, M.~Ben~Amar, S.~Zapperi, and C.~A. La~Porta, ``The role of pressure in cancer growth,'' {\em The European Physical Journal Plus}, vol.~130, p.~224, Nov. 2015.

\bibitem{ben_amar_contour_2011}
M.~Ben~Amar, C.~Chatelain, and P.~Ciarletta, ``Contour {Instabilities} in {Early} {Tumor} {Growth} {Models},'' {\em Physical Review Letters}, vol.~106, p.~148101, Apr. 2011.

\bibitem{ranft_fluidization_2010}
J.~Ranft, M.~Basan, J.~Elgeti, J.-F. Joanny, J.~Prost, and F.~Julicher, ``Fluidization of tissues by cell division and apoptosis,'' {\em Proceedings of the National Academy of Sciences}, vol.~107, pp.~20863--20868, Dec. 2010.

\bibitem{korolev_genetic_2010}
K.~S. Korolev, M.~Avlund, O.~Hallatschek, and D.~R. Nelson, ``Genetic demixing and evolution in linear stepping stone models,'' {\em Reviews of Modern Physics}, vol.~82, pp.~1691--1718, May 2010.

\bibitem{birzu_fluctuations_2018}
G.~Birzu, O.~Hallatschek, and K.~S. Korolev, ``Fluctuations uncover a distinct class of traveling waves,'' {\em Proceedings of the National Academy of Sciences}, vol.~115, pp.~E3645--E3654, Apr. 2018.

\end{thebibliography}

\end{document}